\documentstyle[prl,aps,twocolumn]{revtex}
\begin{document}

\draft
\title{Ground-State Properties of a Rotating Bose-Einstein Condensate 
with Attractive Interaction}

\author{Masahito Ueda$^1$ and Anthony J. Leggett$^2$}

\address{
$^1$Department of Physical Electronics, Hiroshima University,         
Higashi-Hiroshima 739-8527, Japan, and Core Research for Evolutional  
Science and Technology (CREST), JST, Japan \\                         
$^2$Department of Physics, University of Illinois at Urbana-Champaign,
Urbana, IL61801-3080}
\date{\today}

\maketitle

\begin{abstract}
The ground state of a rotating Bose-Einstein condensate with attractive 
interaction in a quasi-one-dimensional torus is studied in terms of   
the ratio $\gamma$ of the mean-field interaction energy per particle to 
the single-particle energy-level spacing. The plateaus of quantized     
circulation are found to appear if and only if $\gamma<1$ with the      
lengths of the plateaus reduced due to hybridization of the condensate  
over different angular-momentum states.                                 
\end{abstract}

The Hess-Fairbank effect~\cite{HF}---disappearance of the angular       
momentum (AM) of liquid helium 4 as it is cooled down to absolute zero  
with its container kept rotating slowly---is an analogue of the Meissner
effect in superconductivity, and it may therefore be regarded as a      
hallmark of superfluidity. The essential requisites for the appearance  
of this effect are the single-valuedness of the wave function and the   
presence of a single Bose-Einstein condensate (BEC). Recent realization 
of BEC of lithium 7~\cite{Bradley} has opened up new possibilities      
associated with the attractive interaction between atoms; here the Fock 
exchange interaction could energetically favor the formation of hybrid  
BECs, which might modify the quantization of circulation and the        
Hess-Fairbank effect. In this Letter we investigate these possibilities 
in terms of the conceptually simple geometry of a quasi-one-dimensional 
torus.                                                                

We consider a system of $N$ weakly interacting bosons with mass $M$,    
confined in an torus of radius $R$ and cross-sectional area             
$S=\pi r^2$, where for simplicity we assume $r\ll R$. This condition    
justifies our assumption that the radial wave function is fixed and     
independent of $\omega$ ---the angular frequency of rotation of the     
torus. At sufficiently low temperature, the interaction between dilute  
hard-core bosons is well approximated by Fermi's contact interaction,   
which is characterized by the s-wave scattering length $a$. The         
associated mean-field interaction energy per particle is given by $gN$, 
where $g=2a\hbar^2/MRS$. The positive (negative) sign of $g$ implies    
that the effective interaction between bosons is repulsive (attractive).
The Hamiltonian of our system in the rotating frame is given, up to     
terms which are constant in our approximation, by                       
\begin{eqnarray}
\hat{H}(\omega)\!&=&\!\sum_l \hbar\omega_{\rm c}\!             
\left(\!l\!-\!\frac{\omega}{2\omega_{\rm c}}\!\right)^2\!\!    
\hat{c}_l^\dagger \hat{c}_l+\frac{g}{2}\!\sum_{l,m,n} \!       
\hat{c}_l^\dagger \hat{c}_m^\dagger \hat{c}_{l+n}\hat{c}_{m-n},
\label{H}     
\end{eqnarray}
where $\omega_{\rm c}=\hbar/2MR^2$ is the critical angular frequency,    
$l,m,$ and $n$ denote the projected angular momenta in units of $\hbar$, 
and $\hat{c}_l^\dagger$ and $\hat{c}_l$ are the creation and annihilation
operators of bosons with AM $l$.                                         
In Eq.~(\ref{H}), we have added the term 
$\sum_l\hbar\omega_{\rm c}(\omega/2\omega_{\rm c})^2\hat{c}_l\hat{c}_l^\dagger
=N\hbar\omega^2/(4\omega_{\rm c})$ which is compensated for by the       
Lagrange multiplier $\alpha$ in Eq.~(\ref{var}) and therefore does not   
modify any result below.

We determine the minimum-energy state of the Hamiltonian~(\ref{H}) within
a Hilbert subspace given by $|\Psi\rangle_{\rm HF}=                      
|\cdots,n_{-l},\cdots,n_{-1},n_0,n_1,\cdots,n_l,\cdots\rangle$, where    
$n_l$ denotes the number of bosons that occupy the state with AM $l$.    
This is nothing but the Hartree-Fock (HF) approximation; other           
possibilities will be discussed later. Because the total number of bosons
is $N$, $n_l$'s should satisfy                                           
\begin{eqnarray}
\sum_{l=-\infty}^\infty n_l=N.
\label{suml}
\end{eqnarray}
The expectation value of the Hamiltonian with respect to the state       
 $|\Psi\rangle_{\rm HF}$ is given by 
\begin{eqnarray}
E(\{n_l\})&=&\sum_l K_l(\omega)
n_l-\frac{g}{2}\sum_ln_l^2+g\left(N^2-\frac{N}{2}\right),
\label{E3}
\end{eqnarray}
where $K_l(\omega)\equiv\hbar\omega_{\rm c}(l-\omega/2\omega_{\rm c})^2$.
The distribution of $\{n_l\}$ is determined so as to minimize            
$E(\{n_l\})$ subject to condition~(\ref{suml}).                          

{\it Case of repulsive interaction.}---
We first show that our ansatz wave function $|\Psi\rangle_{\rm HF}$      
reproduces some well-known results. When $g>0$, it is possible to        
simultaneously minimize the kinetic energy and the interaction energy in 
Eq.~(\ref{E3})                                                           
with $n_l=N$ if $l=[(\omega+\omega_{\rm c})/2\omega_{\rm c}]$ and        
$n_l=0$ otherwise, where the symbol $[x]$ denotes the maximum integer    
that does not exceed $x$. This result implies that a single BEC is       
energetically favorable whether or not it is rotated; Bogoliubov's       
virtual-pair excitations only cause a depletion of the condensate and do 
not alter this conclusion. The single valuedness of the wave function    
dictates that the projected AM be quantized in units of $\hbar$, but one 
needs something more to show that it is quantized in units of $N\hbar$.  
The Onsager-Feynman condition for the quantization of circulation, in    
fact, requires the more stringent latter condition. For the case of      
repulsive interaction, the Fock exchange interaction favors a single     
BEC~\cite{Nozieres}, thereby enforcing sharp transitions between         
different AM states and requiring that the circulation be quantized in a 
uniform system as considered in this Letter. (In a related context,      
Castin and Dum have recently considered the stability of vortices for the
nonuniform case of parabolic potentials~\cite{Castin}. See also          
Refs.~\cite{Wilkin,Butts}.)

{\it Case of attractive interaction.}---
When $g<0$, it is impossible to simultaneously minimize the kinetic     
energy and the interaction energy; the kinetic energy becomes minimal   
when the distribution $\{n_l\}$ peaks sharply at                        
$l=[(\omega+\omega_{\rm c})/2\omega_{\rm c}]$, whereas the interaction  
energy becomes maximal for this case. Were it not for the kinetic term, 
the lowest-energy state would be the one in which the distribution of   
$n_l$ is maximally spread; no single state $l$ could then be            
macroscopically occupied, and there would be no BEC. When the system is 
spatially confined, however, the kinetic term competes with the         
attractive interaction, allowing a metastable condensate to be  formed. 

The minimal-energy distribution $\{n_l\}$ is determined so as to        
minimize $E(\{n_l\})$ in Eq.~(\ref{E3}), subject to condition           
(\ref{suml}), giving         
\begin{eqnarray}
n_l=\frac{N}{\gamma}\left[
\alpha-\left(l-\omega/2\omega_{\rm c}\right)^2\right], 
\label{var}
\end{eqnarray}
where $\alpha$ is a Lagrange multiplier, and $\gamma\equiv|g|N/(\hbar   
\omega_{\rm c})=4N|a|R/S$ is the ratio of the mean-field interaction    
energy per particle to the single-particle energy-level spacing. To find
an estimate of $\gamma$, we rewrite it as                               
$\gamma\sim 4\times10^{-4}N|a|[\AA]R[\mu m]/S[\mu m^2]$. For the case of
lithium 7 with  $|a|\simeq14.6$\AA, $R=1\mu m$, and $r=0.2\mu m$, we 
have $\gamma\sim 0.046N$. With suitable choice of these parameters, it  
is possible to prepare the system both with $\gamma<1$ and with         
$\gamma>1$.     

For $n_l$ to be positive, there must be minimum and maximum values of   
$l$, i.e., $-l_1$ and $l_2$. Equation (\ref{suml}) then becomes         
$\sum_{l=-l_1}^{l_2}n_l=N$, which upon substitution of Eq.~(\ref{var})  
for $n_l$ gives                                                         
\begin{eqnarray}
\alpha\!&=&\!\frac{\gamma}{l_1\!+\!l_2\!+\!1}\!+\!
\tilde{\omega}(\tilde{\omega}\!+\!l_1\!-\!l_2)
+\frac{2(l_1^2\!+\!l_2^2\!-\!l_1l_2)\!+\!l_1\!+\!l_2}{6},
\label{alpha}
\end{eqnarray}
where $\tilde{\omega}\equiv\omega/2\omega_{\rm c}$. With the definitions
of $l_1$ and $l_2$, we have                                             
$(l_1+\tilde{\omega})^2<\alpha\leq(l_1+1+\tilde{\omega})^2$ and         
$(l_2-\tilde{\omega})^2<\alpha\leq (l_2+1-\tilde{\omega})^2$, which lead
to                                                                      
\begin{eqnarray}
& & (l_1\!+\!l_2\!) \ {\rm max}\!\left\{\!
\frac{4l_1\!-\!2l_2\!-\!1}{6}\!+\tilde{\omega},
\frac{4l_2\!-\!2l_1\!-\!1}{6}\!-\!\tilde{\omega}\!\right\}
\!<\!\!\frac{\gamma}{l_1\!+\!l_2\!+\!1} \nonumber\\
& & \leq
(l_1\!+\!l_2\!+\!2)\ {\rm min}\!
\left\{\!\frac{4l_1\!-\!2l_2\!+\!3}{6}\!+\tilde{\omega},
\frac{4l_2\!-\!2l_1\!+\!3}{6}\!-\tilde{\omega}\!\right\}
\label{g1} 
\end{eqnarray}
These inequalities uniquely determine the pair of integers $(l_1,l_2)$  
for a given set of $\gamma$ and $\omega$.

When the torus is at rest (i.e., $\omega=0$), Eq.~(\ref{var}) becomes 
$n_l=(N/\gamma)(\alpha-l^2)$, where $\alpha$ is given from              
Eq.~(\ref{alpha}) with $l_1=l_2$ as                                     
$\alpha=\gamma/(2l_1+1)+l_1(l_1+1)/3$, and Eq.~(\ref{g1}) reduces to    
$l_1(4l_1^2-1)/3<\gamma \leq(l_1+1)[4(l_1+1)^2-1]/3$. These inequalities
uniquely determine the number $2l_1+1$ of macroscopically occupied AM   
states for a given $\gamma$. For example, $\gamma\leq 1$,               
$1<\gamma\leq 10$ and  $10<\gamma\leq 35$ correspond to $2l_1+1= 1$, 3, 
and 5, respectively. Thus, there is a single BEC when $\gamma\leq1$.    
This condition agrees with the usual criterion for a metastable BEC to  
exist that is obtained for a parabolically confining potential using the
Gross-Pitaevskii (GP) equation~\cite{Ruprecht}. A new finding in our    
analysis is that for $\gamma>1$ BEC becomes hybridized over different AM
states.                                                                 

At the continuum limit $\omega_{\rm c}\rightarrow0$ (i.e., $R\rightarrow
\infty$) with $|g|N$ held constant, $\gamma$ and $l_1$ become infinite  
with $\alpha/\gamma\sim O (1/l_1)$. It follows from the relation        
$n_l=(N/\gamma)(\alpha-l^2)$ that all $n_l's$ becomes vanishingly small,
of the order of $N/l_1$. Thus, no BEC exists for an infinite system in  
accordance with the standard widsom~\cite{Nozieres}.

The analysis for the case of $\omega\neq0$ is straightforward, and we   
describe here only the results that are relevant to later discussions.  

\noindent
1. The region in which a single BEC with $n_l=N$ exists is given from   
Eq.~(\ref{g1}) with $l_1=-l,l_2=l$ by
\begin{eqnarray}
0<\gamma \leq
-|\frac{\omega}{\omega_{\rm c}}-2l|+1.
\label{g11}   
\end{eqnarray}
When $\omega/\omega_{\rm c}$ is an odd integer, condition~(\ref{g11})   
can never be met, so no unique BEC can exist no matter how weak the     
attractive interaction.                                                 

\noindent
2. The region in which two states with AM $l$ and $l+1$ are             
macroscopically occupied is given from  Eq.~(\ref{g1}) with             
$l_1=-l,l_2=l+1$ by
\begin{eqnarray}
|\frac{\omega}{\omega_{\rm c}}\!-\!2l\!-\!1|\!<\!\gamma\!\leq\!
{\rm min}\left\{\!
3\frac{\omega}{\omega_{\rm c}}\!-\!6l\!+\!1,
-3\frac{\omega}{\omega_{\rm c}}\!+\!6l\!+\!7\right\},
\label{g12}   
\end{eqnarray}
and the corresponding distribution of bosons is given by
\begin{eqnarray}
n_l=\frac{N}{2}\left[1-
\frac{\omega-(2l+1)\omega_{\rm c}}{\gamma\omega_{\rm c}}
\right], \ \ \ n_{l+1}=N-n_l.
\label{binary}
\end{eqnarray}

\noindent
3. In general, the region in which $k$ states with AM                   
$l,l+1,\cdots,l+k-1$ are macroscopically occupied is given from         
Eq.~(\ref{g1}) with $l_1=-l,l_2=l+k-1$ by
\begin{eqnarray}
& & k(k-1){\rm max}\!\left\{\!
\frac{-6l\!-\!2k\!+\!1}{6}\!+\!\frac{\omega}{2\omega_{\rm c}}, 
\frac{6l\!+\!4k\!-\!5}{6}\!-\!\frac{\omega}{2\omega_{\rm c}}\!\right\}
\!<\gamma \nonumber\\
& & \leq\!    
k(k\!+\!1) {\rm min}\!\left\{\!
\frac{-\!6l\!-\!2k\!+\!5}{6}\!+\!\frac{\omega}{2\omega_{\rm c}},         
\frac{6l\!+\!4k\!-\!1}{6}\!-\!\frac{\omega}{2\omega_{\rm c}}\!\right\}\!.
\label{g1k}   
\end{eqnarray}
The phase diagram is shown in Fig.~\ref{fig:PD}. We have thus shown      
that there are regions of $\gamma$ and $\omega$ in  which more than one  
AM state is macroscopically occupied. This prediction can be tested most 
directly by switching off the trap potential and let the system expand.  
Due to Heisenberg's uncertainty relation, the tight radial confinement of
the trap causes the gas to expand more rapidly in that direction than in 
other ones, and the superposition of BECs having different AM should     
result in an interference pattern with broken axisymmetry~\cite{Mueller}.

{\it Partial quantization of circulation.}---
When we fix $\gamma\equiv|g|N/\hbar\omega_{\rm c}<1$ and increase        
$\omega$ from $0$, we alternatively pass regions in which one or two AM  
states are macroscopically occupied (see Fig~\ref{fig:PD}), and in the   
latter regions the distribution of bosons between the two AM states      
changes continuously with $\omega$, as can be seen from                  
Eq.~(\ref{binary}). What happens then to the circulation $\kappa$ of the 
system? When a single BEC with AM $l$ exists, $\kappa$ is given by $hl/M$. 
When two AM states are macroscopically occupied, $\kappa$ should be given
by $h\langle l\rangle/M$, where $\langle l\rangle$ is the
ensemble-averaged value of the AM. To find this value, let us restrict   
ourselves to the region 
\begin{eqnarray}
|(\omega\!-\!\omega_{\rm c})/\omega_{\rm c}|<\gamma\leq 
-3|(\omega-\omega_{\rm c})/\omega_{\rm c}|\!+\!4,
\label{cond3}
\end{eqnarray}
where two states with AM $l=0$ and $l=1$ are macroscopically occupied,   
and the number of bosons in each condensate is given from                
Eq.~(\ref{binary}) by                                                    
$n_0=N/2-N(\omega-\omega_{\rm c})/(2\omega_{\rm c}\gamma)$ and           
$n_1=N-n_0$. Hence the ensemble-averaged AM $\langle l\rangle$ is given by
\begin{eqnarray}
\langle l\rangle
=\frac{1}{2}+\frac{\omega-\omega_{\rm c}}{\omega_{\rm c}}
\frac{\hbar\omega_{\rm c}}{2|g|N}
=\frac{1}{2}+\frac{\omega-\omega_{\rm c}}{\omega_{\rm c}}
\frac{S}{8N|a|R},
\label{av}
\end{eqnarray}
which does not show any sharp transition (see  Fig.~\ref{fig:AM}), in    
sharp contrast with the case of repulsive interaction. Suppose now       
that we perform the Hess-Fairbank experiment for the frequency of        
rotation and for the strength of interaction that satisfy the condition  
(\ref{cond3}). Then the AM will not completely be ^^ ^^ expelled"        
even at absolute zero and  have a nonzero value given by~Eq.~(\ref{av}). 
Only when those parameters are in the region (\ref{g11}) with $l=0$, the 
AM should vanish at absolute zero.

{\it Hybrid BECs vs. a phase-coherent single BEC.}---                
We have shown within the HF approximation that hybrid BECs exist for 
some ranges of parameters $\gamma$ and $\omega/\omega_{\rm c}$.      
Recently, Rokhsar has argued that hybrid or ^^ ^^ fragmented" BECs   
are inherently unstable against the formation of a single BEC whose  
macroscopically occupied state is a linear combination of the        
^^ ^^ fragments" with definite relative phases~\cite{Rokhsar}. In our
situation, the ^^ ^^ fragments" refer to macroscopically             
occupied AM states. We first discuss the stability of binary BECs    
against forming such a phase-coherent single BEC. Because the        
properties of the system are periodic functions of $\omega$ with     
periodicity $2\omega_{\rm c}$, we may consider, without loss of      
generality, the region (\ref{cond3}) in which two BECs with $l=0$ and
$l=1$ coexist. The state vector of this binary BECs is given by      
\begin{eqnarray}
|\Psi\rangle_{\rm HF}=|n_0,n_1\rangle=\frac{1}{\sqrt{n_0!n_1!}}      
(\hat{c}_0^\dagger)^{n_0}(\hat{c}_1^\dagger)^{n_1}|{\rm vac}\rangle, 
\label{HF}
\end{eqnarray}
where $n_0$ and $n_1$ are the numbers of bosons in the $l=0$ and     
$l=1$ states, which are given by Eq.~(\ref{binary}). To be compared  
with this state is a single macroscopically occupied state whose     
creation operator $\hat{b}^\dagger$ is given by                      
$\hat{b}^\dagger=\alpha\hat{c}_0^\dagger+\beta\hat{c}_1^\dagger$,    
where $\alpha$ and $\beta$ are determined so as to minimize the total
energy, subject to $|\alpha|^2+|\beta|^2=1$. The corersponding single
BEC is given by                                                      
\begin{eqnarray}
|\Psi\rangle_{\rm single}=\frac{(\hat{b}^\dagger)^N}{\sqrt{N!}}      
|{\rm vac}\rangle=\frac{1}{\sqrt{N!}}                                
(\alpha\hat{c}_0^\dagger+\beta\hat{c}_1^\dagger)^N|{\rm vac}\rangle. 
\label{single}
\end{eqnarray}
The crucial observation here is that when only two states are macroscopically
occupied, the expectation value $\langle\hat{H}\rangle_{\rm single}$ of
the Hamiltonian~(\ref{H}) over the state (\ref{single}) does not       
contain any non-HF terms that are of the same order of magnitude as the
HF terms because of the conservation of AM. Therefore, the system      
cannot lower its energy by establishing a relative phase coherence     
between the different AM states. The minimum value of                  
$\langle\hat{H}\rangle_{\rm single}$ is reached when
\begin{eqnarray}
|\alpha|^2=\frac{1}{2}\left[1-
\frac{\omega-\omega_{\rm c}}{\gamma\omega_{\rm c}}\frac{N}{N-1}\right],
\ \ |\beta|^2=1-|\alpha|^2,
\end{eqnarray}
and by a straightforward calculation, we find that
\begin{eqnarray}
\langle\hat{H}\rangle_{\rm HF}-\langle\hat{H}\rangle_{\rm single}
\simeq\frac{|g|N}{4}\left[\left(
\frac{\omega-\omega_{\rm c}}{\gamma\omega_{\rm c}}\right)^2-1\right]<0. 
\label{dif}
\end{eqnarray}
However, because the energy difference is of the order of $1/N$, the  
two states (\ref{HF}) and (\ref{single}) are virtually degenerate. In 
real life, 
however, there are inhomogeneities in the container ^^ ^^ walls" etc.,
which break the exact axisymmetry.                                   
Such a perturbation, however weak, could stabilize the single         
coherent BEC relative to a ^^ ^^ fragmented" one. To show this,       
consider a symmetry-breaking perturbation that  mixes the $l=0$ state 
and the $l=1$ state:                                                  
$\hat{V}=t\hat{c}_0^\dagger\hat{c}_1+t^*\hat{c}_1^\dagger\hat{c}_0$.  
It is easy to see that while $\hat{V}$ does not lower the energy of   
the system for the HF state ($\langle\hat{V}\rangle_{\rm HF}=0$), it  
does for the single coherent BEC;                                     
$\langle\hat{V}\rangle_{\rm single}=2N{\rm Re}(\alpha^*\beta t)       
=-2N|\alpha^*\beta t|$, provided that arg$\alpha$-arg$\beta$-arg$t=\pm\pi$.
Because both $l=0$ and $l=1$ states are macroscopically occupied, i.e.
$\alpha\sim O(1)$ and $\beta\sim O(1)$,                               
$\langle\hat{V}\rangle_{\rm single}$ is extensive. The single coherent
BEC can therefore become energetically favorable due to a (possibly   
infinitesimal) symmetry-breaking perturbation. It should be noted,    
however, that 
the plot of $\langle l\rangle$ versus $\omega$ in Fig.~\ref{fig:AM}   
remains basically unaltered because it does not depend on whether or  
not a phase coherence is established between two macroscopically      
occupied AM state.                                                    

The situation is different when more than two AM states are           
macroscopically occupied. Now the expectation value of the Hamiltonian
contains non-HF terms that are of the same order of magnitude as the  
HF terms, so that without the need of the symmetry-breaking           
perturbation the system can lower its energy by establishing a        
relative phase coherence. To show this, let us consider the case of   
$1+3|\omega/\omega_{\rm c}|<\gamma<10-6|\omega/\omega_{\rm c}|$, where
three AM states $l=-1,0,1$ are macroscopically occupied. The hybrid BEC
state is described by $|\Psi\rangle_{\rm HF}=|n_{-1},n_0,n_1\rangle$  
and the corresponding single coherent BEC is described by             
$|\Psi\rangle_{\rm single}=(\alpha\hat{c}_{-1}^\dagger+\beta          
\hat{c}_0^\dagger+\gamma\hat{c}_1^\dagger)^N/\sqrt{N!}|{\rm vac}      
\rangle$ with $|\alpha|^2+|\beta|^2+|\gamma|^2=1$. The expectation    
value of the Hamiltonian with respect to $|\Psi\rangle_{\rm HF}$ is   
given by 
\begin{eqnarray}
& & \langle\hat{H}\rangle_{\rm HF}=K_{-1}n_{-1}+K_0n_0+K_1n_1\nonumber\\
& & -|g|(n_{-1}n_0+n_0n_1+n_1n_{-1})\!-|g|N(N\!-\!1)/2,
\end{eqnarray}
which is minimized when
$n_{\mp 1}=N[1\mp(3\omega\pm\omega_{\rm c})/(\gamma\omega_{\rm c})]/3$
and $n_0=N[1+2/\gamma]/3$.                                            
The expectation value of the Hamiltonian with respect to              
$|\Psi\rangle_{\rm single}$ is given by                               
\begin{eqnarray}
& & \langle\hat{H}\rangle_{\rm single}=N(K_{-1}|\alpha|^2+K_0|\beta|^2+  
K_1|\gamma|^2) \nonumber\\
& & -|g|N(N-1)(|\alpha|^2|\beta|^2+|\beta|^2|\gamma|^2+       
|\gamma|^2|\alpha|^2)\nonumber\\
& & -|g|N(N-1)/2
-|g|N(N-1)(\alpha\beta^{*2}\gamma+\alpha^*\beta^2\gamma^*).             
\end{eqnarray}
Because the last two terms are phase-dependent and of the same order  
of magnitude as the remaining terms, it is clear that the single      
coherent BEC can have a lower energy than the fragmented BEC state by,    
e.g., the following choice of amplitudes,           
$\alpha=\sqrt{n_{-1}}e^{i\theta_\alpha}, \
\beta =\sqrt{n_0}   e^{i\theta_\beta},  \ 
\gamma=\sqrt{n_1}   e^{i\theta_\gamma}$   
with the relative phase relation
$\theta_\alpha-2\theta_\beta+\theta_\gamma=0.$ Thus the ternary 
BEC state is unstable against the formation of a single coherent BEC. 
Similar mechanisms should work when more than three AM states are     
macroscopically occupied.                                             

To summarize, we have studied the ground-state properties of a        
rotating BEC with attractive interaction confined in a                
quasi-one-dimensional torus. When the condition (\ref{g11}) is met,   
only one AM state is macroscopically occupied. When the condition     
(\ref{g12}) is met, two BECs with different AM can, in principle,     
coexist. However, any deviation from the exact axisymmetry            
is shown to stabilize a single coherent BEC relative to a             
^^ ^^ fragmented" one. The plateaus of quantized circulation appear if
$\gamma<1$, but the lengths of the plateaus are reduced. In other     
regions of parameters $\gamma$ and $\omega$, more than two AM states  
are macroscopically occupied, where non-HF terms stabilize a single   
coherent BEC even in the presence of the exact axisymmetry.

This work was supported in part by the National Science Foundation    
under grant no. DMR-96-14133 and by a Grant-in-Aid for Scientific     
Research (Grant No. 08247105) by the Ministry of Education, Science,  
Sports, and Culture of Japan.


\begin{figure}
\caption{Regions of $\omega/\omega_{\rm c}$ and 
$\gamma\equiv |g|N/\hbar\omega_{\rm c}$ showing various phases of 
coexisting macroscopically occupied angular-momentum (AM) states. 
The triangles show regions of single Bose-Einstein condensates    
(BECs) with AM $l=-3,-2,\cdots,3$ from left to right. The diamond 
indicated by $(-3,-2)$, for example, shows the region in which two
AM states $l=-3$ and $-2$ are macroscopically occupied. }
\label{fig:PD}

\medskip
\caption{Ensemble-averaged projected angular momentum 
$\langle l \rangle$ in units of $\hbar$ as a function 
of $\omega/\omega_{\rm c}$ for $\gamma\leq 1$. The circulation
is given by $\kappa=h\langle l\rangle/M$.
The crossover regions between plateaus correspond to the  
regions in which two AM states are macroscopically occupied. 
When $\gamma$ exceeds one, the plateaus disappear.}
\label{fig:AM}
\end{figure}

\end{document}